\documentstyle[twocolumn,aps,prl,floats,epsf,twoside]{revtex}
\begin{document} 
\twocolumn[\hsize\textwidth\columnwidth\hsize\csname
@twocolumnfalse\endcsname
\draft
\title{The Dynamics of Curved Fronts: Beyond Geometry} 

\author{Aric Hagberg \cite{byline1}}
\address{Center for Nonlinear Studies and T-7, Theoretical Division,
Los Alamos National Laboratory, Los Alamos, NM 87545}

\author{Ehud Meron \cite{byline2}}
\address{The Jacob Blaustein Institute for Desert Research and the Physics 
Department, Ben-Gurion University, \\ Sede Boker Campus 84990, Israel}

\date{\today}

\maketitle

\begin{abstract}
We derive a new set of kinematic equations for front motion in
two-dimensional bistable media.  The equations generalize the
geometric approach by complementing the equation for the front
curvature with an order parameter equation associated with a
nonequilibrium Ising-Bloch bifurcation. The resulting equations
capture the core structure of spiral waves and spontaneous spiral-wave
nucleation.
\end{abstract}

\pacs{PACS numbers: 47.20.Ma, 82.20.Mj, 82.40.Ck}
\vspace{-0.8cm}
\vskip2pc]

\narrowtext

Traveling wave phenomena in reaction-diffusion systems often involve
sharp interfaces or fronts separating different reaction states. The
dynamics of two-dimensional sharp fronts has been studied successfully
using a geometric
approach~\cite{Zykov:87,Mikhailov:90,MePe:88,Meron:92,Brazhnik:96}.
Given a relation between the normal velocity of the front and its
curvature, the geometric theory consists of a closed
integro-differential equation for the front curvature from which the
front line shape in the physical plane can be extracted. Inherent in
this approach is the assumption that the inner front structure does
not change significantly in time.  This assumption rules out major
changes like spontaneous nucleation of spiral waves along the front.
Such phenomena have been observed recently in numerical simulations of
model equations describing bistable reaction-diffusion systems. Very
often the nucleation of spiral waves triggers spot replication and
spiral turbulence~\cite{LeSw:95,HaMe:94b,HaMe:94c,EHM:95}.

In this Letter we present a new kinematic approach for front motion in
two-dimensional bistable media that captures spontaneous spiral-wave
nucleation along the front.  A key step in this approach is the
consideration of a parameter range including a nonequilibrium
Ising-Bloch (NIB) front bifurcation. This parity breaking bifurcation
renders a stationary planar front unstable and gives rise to a pair of
stable counter-propagating fronts. The bifurcation has been found in a
number of models, including the forced complex
Ginzburg-Landau~\cite{CLHL:90} and
FitzHugh-Nagumo~\cite{IMN:89,HaMe:94a,BRSP:94} equations, and in
experiments with chemical reactions~\cite{HBKR:95} and liquid
crystals~\cite{NYK:95}.

Our kinematic approach consists of three equations:
\begin{itemize}
\item A geometric equation for the front curvature, $\kappa$:
\begin{equation}
{\partial\kappa\over\partial t} = -(\kappa^2 + {\partial^2\over\partial
s^2})C_n - {\partial\kappa\over\partial s}\int_0^s \kappa C_n ds^\prime \,.
\label{K}
\end{equation}
\item An equation  relating the normal front velocity $C_n$, the curvature
$\kappa$, and the order parameter, $C_0$, associated with the NIB 
bifurcation:
\begin{equation}
C_n = C_0 - D\kappa\,.
\label{Cn}
\end{equation}
\item An equation for the order parameter:
\begin{eqnarray}
{\partial C_0\over\partial t}&=&(\alpha_c-\alpha)C_0 - \beta C_0^3
+\gamma\kappa + \gamma_0 \nonumber \\
&&\mbox{}
+{\partial^2 C_0\over \partial s^2} 
- {\partial C_0 \over \partial s} \int_0^s \kappa C_n ds^\prime\,. 
\label{C0}
\end{eqnarray}
\end{itemize}
In these equations $s$ is the front arclength, and the critical parameter
value $\alpha_c$ designates the NIB bifurcation point. Notice that $C_0$
coincides with the planar front velocity when $\kappa=0$.

The curvature equation~(\ref{K}) together with the eikonal
equation~(\ref{Cn}), where $C_0$ is considered {\em constant},
constitute the geometric approach used in earlier
studies~\cite{Mikhailov:90,Meron:92}.  Relaxing the requirement of
constant $C_0$ by adding Eq.~(\ref{C0}) allows for spontaneous local
reversal of the direction of front propagation.  The reversals are
accompanied by the nucleation of spiral-wave pairs. In the rest of
this Letter we describe the derivation of Eqs.~(\ref{Cn})
and~(\ref{C0}) for a particular model and use these equations to
demonstrate a mechanism of spontaneous spiral-wave nucleation.

We consider the FitzHugh-Nagumo model with a diffusing inhibitor,
\begin{eqnarray}
{\partial u\over\partial t}&=& 
\epsilon^{-1}(u-u^3-v)+\delta^{-1}\nabla^2u\,,\nonumber \\
{\partial v\over\partial t}&=& u-a_1v-a_0+\nabla^2 v\,, \label{FHN}
\end{eqnarray} 
where $u$ and $v$, the activator and the inhibitor, are real scalar fields
and $\nabla^2$ is the Laplacian operator in two dimensions. The parameter
$a_1$ is chosen so that~(\ref{FHN}) describes a bistable medium having two
stable uniform states: an ``up'' state $(u_+,v_+)$ and a ``down'' state
$(u_-,v_-)$. 
Ising and Bloch front solutions connect the two uniform states
$(u_\pm,v_\pm)$ as the spatial coordinate normal to the front goes from 
$-\infty$ to $+\infty$.
The parameter space of interest is spanned by $\epsilon,
\delta$ and $a_0$, or alternatively by 
$\eta=\sqrt{\epsilon\delta}$, $\mu=\epsilon/\delta$,
and $a_0$. Note the parity symmetry
$(u,v)\to(-u,-v)$ of~(\ref{FHN}) for $a_0=0$. 

\begin{figure}
\epsfxsize=3.0 truein \epsffile{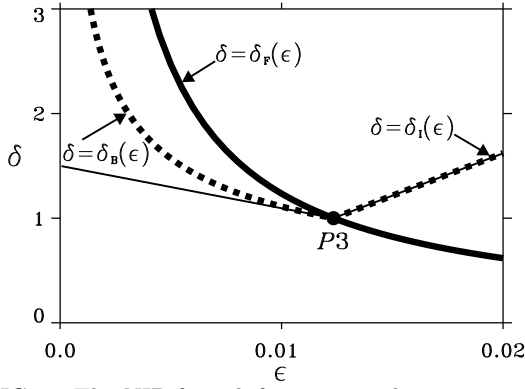}
\caption{The NIB front bifurcation and transverse instability
boundaries.  The thick line is the NIB bifurcation, $\delta(\epsilon)$,
and the dashed lines are the boundaries for transverse instability
of Ising, $\delta_I(\epsilon)$, and Bloch, $\delta(\epsilon)$, fronts.
The thin lines are the linear approximations to the transverse
instability boundaries near the codimension 3 point, $P3$. 
Parameters: $a_1=4.0$, $a_0=0$.}
\label{fig:lines}

\end{figure}

The NIB bifurcation line for $a_0=0$ is shown in
Fig.~\ref{fig:lines}. For $\mu\ll 1$ it is given by
$\delta=\delta_F(\epsilon)=\eta_c^2/\epsilon$, or $\eta=\eta_c$, where
$\eta_c=\frac{3}{2\sqrt{2}q^3}$ and $q^2=a_1+1/2$~\cite{HaMe:94a}.
The single stationary Ising front that exists for $\eta>\eta_c$ loses
stability to a pair of counter-propagating Bloch fronts at
$\eta=\eta_c$.  Beyond the bifurcation ($\eta<\eta_c$) a Bloch front
pertaining to an up state invading a down state coexists with another
Bloch front pertaining to a down state invading an up state. Also
shown in Fig.~\ref{fig:lines} are the transverse instability
boundaries (for $a_0=0$), $\delta=\delta_I(\epsilon)=
\epsilon/\eta_c^2$ and $\delta=\delta_B(\epsilon)=\eta_c/\sqrt\epsilon$,
for Ising and Bloch fronts respectively.  Above these lines,
$\delta>\delta_{I,B}$, planar fronts are unstable to transverse
perturbations~\cite{HaMe:94b,HaMe:94c}.  All three lines meet at a
codimension 3 point $P3$: $\epsilon=\eta_c^2$, $\delta=1$, $a_0=0$.

The following assumptions are made to derive Eqs.~(\ref{Cn})
and~(\ref{C0}): $\epsilon$ and $\delta$ are in the proximity of
the codimension 3 point, $P3$, with $\epsilon<<1$; the radius
of curvature is much larger than the front width, that is,
$\kappa\ll q$.  First, we transform to an orthogonal coordinate
system $(r,s)$ that moves with the front, where $r$ is a
coordinate normal to the front.  Let ${\bf X}(s,t)=(X,Y)$
denote the position vector of the front.  More precisely we
identify ${\bf X}(s,t)$ with the $u=0$ contour line.  The
relation between the laboratory frame $(x,y)$ and the moving
frame is
\begin{equation}
x=X(s,t)+r{\partial Y\over\partial s}\qquad 
y=Y(s,t)-r{\partial X\over\partial s}\,.
\label{trans}
\end{equation}
In the moving frame Eqs.~(\ref{FHN}) become
\begin{eqnarray}
\epsilon{\cal D}u&=& 
u-u^3-v + \mu{\cal L}u\,,\nonumber \\
{\cal D}v&=& u-a_1v-a_0+ {\cal L}v\,, \label{mFHN}
\end{eqnarray} 
where
\begin{eqnarray}
{\cal D}&=&{\partial\over\partial t} + {\partial s\over\partial
t}{\partial\over\partial s} + {\partial r\over\partial
t}{\partial\over\partial r}\,,\nonumber \\
{\cal L}&=&{\partial^2\over\partial r^2} + \kappa F{\partial\over\partial r}
+ F{\partial F\over\partial s}{\partial\over\partial s} +
F^2{\partial^2\over\partial s^2}\,, \nonumber
\end{eqnarray}
$F=(1+r\kappa)^{-1}$, and the curvature is $\kappa={\partial X\over\partial
s}{\partial^2 Y\over\partial s^2}-{\partial Y\over\partial
s}{\partial^2 X\over\partial s^2}$~\cite{Keener:86}.
The change of the arclength in time is due to stretching and is 
given by~\cite{Mikhailov:90,Meron:92}
\begin{equation}
{\partial s\over\partial t}=\int_0^s \kappa C_n ds^\prime \,.
\label{st}
\end{equation}

Recalling that $\mu=\epsilon/\delta\ll 1$, we use singular
perturbation theory and distinguish between an inner region where
$\mu\frac{\partial^2 u}{\partial r^2}\sim {\cal O}(1)$, and outer
regions where $\mu{\partial^2 u\over\partial r^2}\sim {\cal O}(\mu)$.
The inner region pertains to the front core where the profile of $u$
in the normal direction is steep.  Introducing a stretched coordinate
$z=r/\sqrt{\mu}$ and expanding $u=u_0+\epsilon u_1 + \epsilon^2 u_2
+\ldots$ and $v=v_0+\epsilon v_1 + \epsilon^2 v_2 +\ldots\,$, we
obtain at order unity $u_0=-\tanh z/\sqrt 2$, $v_0=0$.  At order
$\epsilon$ the solvability condition is
\begin{equation}
{\partial r\over\partial t}={3\over\eta\sqrt 2}v_f +
\delta^{-1}\kappa\ ,
\label{rt}
\end{equation}
where $v_f=v(0,s,t)+{\cal O}(\epsilon^2)$ is the approximately
constant value of the inhibitor $v$ in the narrow [${\cal
O}(\sqrt\mu)$] front core region. The first term on the
right-hand-side of~(\ref{rt}) is identified with the order parameter
for the NIB bifurcation: $C_0=-{3\over\eta\sqrt 2}v_f$.  Since the
normal velocity is $C_n=-{\partial r\over\partial t}$, Eq.~(\ref{rt})
yields the eikonal equation~(\ref{Cn}) with $D=\delta^{-1}$.

In the outer regions ${\partial^2 u\over\partial r^2} \sim {\cal O}(1)$ and
the leading order equation for $u$ is $u-u^3-v=0$. The relevant solutions
are $u=u_+(v)\approx 1-v/2$ for $r<0$ and $u=u_-(v)\approx -1-v/2$ for
$r>0$ (assuming $a_1$ is sufficiently large)~\cite{HaMe:94a}. To leading 
order in
$\epsilon$ we obtain for $v$ the free boundary problem
\begin{eqnarray}
\left({\partial \over\partial t}-{\partial^2\over\partial r^2}+q^2\right)v
&=&+1 -{3\over \eta\sqrt 2}v_f{\partial v\over\partial r}+P_1+P_2,
\quad r\le 0 \nonumber \\
\left({\partial \over\partial t}-{\partial^2\over\partial r^2}+q^2\right)v
&=&-1 -{3\over \eta\sqrt 2}v_f{\partial v\over\partial r}+P_1+P_2,
\quad r\ge 0 \nonumber \\
v(\mp\infty,s,t)&=&v_\pm={\pm 1 - a_0\over q^2}\ ,\nonumber \\
\left[v\right]_{r=0}&=&\left[{\partial v\over\partial r}\right]_{r=0}=0\ ,
\label{free}
\end{eqnarray}
where
\begin{eqnarray}
P_1&=&(1-\delta^{-1})\kappa{\partial v\over\partial r}-a_0+F^2{\partial^2
v\over \partial s^2}-{\partial s\over\partial t}
{\partial v\over\partial s}\,, 
\\
P_2&=&F{\partial F\over\partial s}{\partial v\over\partial s}\, ,
\end{eqnarray}
and the square brackets denote jumps of the quantities inside the brackets
across the front at $r=0$.

To solve this free boundary problem we consider a parameter range in
the immediate vicinity of the $P3$ point in Fig.~\ref{fig:lines}. In
that range the transverse instabilities of the fronts involve only
small wavenumbers and therefore we can assume weak dependence of $v$
and $\kappa$ on the arclength $s$. In addition, the front speed is
small and vanishes at $P3$. This suggests using the speed of a planar
Bloch front solution, $c\propto\sqrt{\eta_c-\eta}$, as a small
parameter. The weak dependence of $v$ and $\kappa$ on $s$ is achieved
by introducing the slow length scale $S=cs$ and assuming ${\bf X}=
{\bf X}(S,t)$.  This assumption dictates $\kappa=c^3\kappa_0$ where
$\kappa_0\sim {\cal O}(1)$. We also introduce a slow time scale $T=c^2
t$ to describe deviations from steady front motion.

Following Ref.~\cite{HMRZ:96}, we solve the free boundary problem~(\ref{free}) 
by expanding propagating curved front solutions as power
series in $c$ around the stationary planar Ising front
\begin{equation}
v(r,S,t,T)=v^{(0)}(r)+\sum_{n=1}^\infty c^nv^{(n)}(r,S,t,T)\ ,
\label{vrStT}
\end{equation}
where $v^{(0)}(r)=(1-e^{qr})/q^2$ for $r\le 0$ and 
$v^{(0)}(r)=(e^{-qr}-1)/q^2$ for $r\ge 0$. Expanding 
$\eta=\eta_c-c^2\eta_1+c^4\eta_2+...$ and
using these expansions in 
(\ref{free}) produces the set of equations
\begin{equation}
\frac{\partial v^{(n)}}{\partial t} + q^2 v^{(n)} - 
\frac{\partial^2 v^{(n)}}{\partial r^2} =
 -\rho^{(n)},~n=1,2,3,\ldots
\label{vn}
\end{equation}
where $\rho^{(1)}$ and $\rho^{(2)}$ are as in Ref. \cite{HMRZ:96} 
and $\rho^{(3)}$ is 
\begin{eqnarray}
\lefteqn{\rho^{(3)}(r,S,t,T)={\partial v^{(1)}\over\partial T} 
+ {3\eta_1\over \sqrt 2 \eta_c^2} v^{(1)}_{\vert r=0}
{\partial v^{(0)}\over\partial r}}\hspace{0.4cm} && \nonumber\\
&&\mbox{} + {3\over\sqrt 2 \eta_c}\left[v^{(1)}_{\vert r=0}{\partial 
v^{(2)}\over\partial r} + v^{(2)}_{\vert r=0}{\partial v^{(1)}\over 
\partial r} + v^{(3)}_{\vert r=0}{\partial v^{(0)}\over\partial r}\right]
+ a_{00} \nonumber \\
&&\mbox{}- F^2{\partial^2 v^{(1)}\over\partial S^2} 
-(1-\delta^{-1})\kappa_0{\partial v^{(0)}\over\partial r} +  
{\partial S\over\partial T}{\partial v^{(1)}\over\partial S} \ .
\label{rho3}
\end{eqnarray}
In (\ref{rho3}) we assumed $a_0=c^3 a_{00}$ where $a_{00}\sim {\cal 
O}(1)$, and recall that $\kappa_0=\kappa/c^3$. Notice that ${\partial
S\over\partial T}\sim{\cal O}(1)$, and $P_2$ contributes only at orders 
higher than $c^3$.
We solve Eq.~(\ref{vn}) using the asymptotic behavior of
an appropriate Green's function as in Ref.~\cite{HMRZ:96}. The results for 
$n=1,2$ 
remain unchanged and give the front bifurcation point
$\eta_c={3\over2\sqrt 2 q^3}$. 
The solution of~(\ref{vn}) with $n=3$ yields the compatibility condition
\begin{eqnarray}
{\partial v^{(1)}\over\partial T}& =& {\sqrt 2 \eta_1\over q\eta_c^2}v^{(1)}
- {3\over 4\eta_c^2} {v^{(1)}}^3 - {4\over 3}a_{00} \nonumber \\
&&\mbox{}- {2(1-\delta^{-1})\over 3q}\kappa_0 + {\partial^2 v^{(1)}\over\partial 
S^2} - {\partial S\over\partial T}{\partial v^{(1)}\over\partial S}\ ,
\label{comp}
\end{eqnarray}
or expressing the slow time and arclength derivatives in terms of the 
fast variables $t,s$ and using (\ref{st}) and (\ref{vrStT}),
\begin{eqnarray}
{\partial v_f\over\partial t}& =& {\sqrt 2 (\eta_c-\eta)\over q\eta_c^2}v_f 
-{3\over 4\eta_c^2}v_f^3 - {4\over 3}a_0 \nonumber \\
&&\mbox{} -{2(1-\delta^{-1})\over 3q}\kappa
+ {\partial^2 v_f\over \partial s^2}-{\partial v_f\over\partial s}
\int_0^s \kappa C_n ds^\prime \ .
\label{vfeqn}
\end{eqnarray}
Equation~(\ref{vfeqn}) coincides with~(\ref{C0}) once we make the 
following identifications: $C_0=-{3\over\eta\sqrt 2}v_f$, 
$\alpha={\eta\sqrt 2\over q\eta_c^2}$, $\alpha_c={\sqrt 2\over q\eta_c}$, 
$\beta=1/6$, $\gamma=\alpha_c(1-\delta^{-1})$,  and $\gamma_0=2\alpha_cqa_0$.

Equation (\ref{C0}) reproduces the NIB bifurcation for planar fronts:
setting $\kappa=0$ and $a_0=0$ we find the Ising front branch $C_0=0$
and the two Bloch front branches
$C_0=\pm\sqrt{(\alpha_c-\alpha)/\beta}$.  To test whether
Eqs.~(\ref{K})-(\ref{C0}) also capture the transverse instabilities we
check the linear stability of planar front solutions near the $P3$
point in the $a_0=0$ plane.  Let $C_0=C_0^0 + \bar C_0\exp(\sigma
t+iQs)$ and $\kappa=\kappa^0 + \bar\kappa\exp(\sigma t+iQs)$ where
$(C_0^0,\kappa^0)=(0,0)$ for the Ising front and
$(C_0^0,\kappa^0)=(\pm\sqrt{(\alpha_c-\alpha)/\beta},0)$ for the Bloch
fronts.  Inserting these forms in~(\ref{C0}) gives the following
transverse instability lines, linearized around $\delta=1$:
\[
{\rm Ising:} ~~~\epsilon=\eta_c^2\delta\,, \qquad {\rm Bloch:}
~~\epsilon=\eta_c^2(3-2\delta) \,.
\]
These lines are displayed in Fig.~\ref{fig:lines} (thin lines). To
linear order around the $P3$ point they coincide with the exact
transverse instability lines.

\begin{figure}[htb]
\epsfxsize=3.0in \epsffile{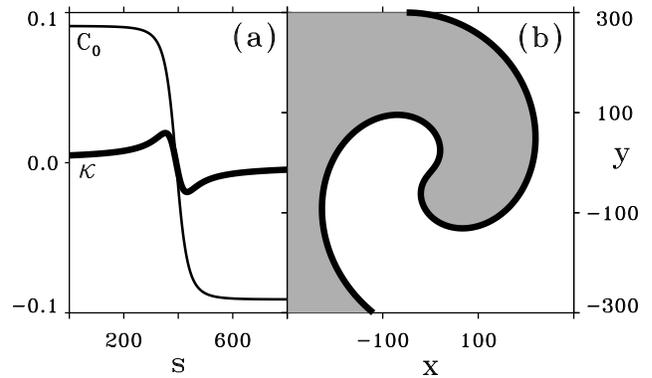}
\caption{A front solution to the kinematic equations 
(\protect\ref{K})-(\protect\ref{C0}).
(a) The order parameter $C_0$ and the curvature $\kappa$
along the arclength $s$.
(b) In the $x-y$ plane the front solution corresponds to a rotating
spiral wave. The shaded (light) region corresponds to an up (down) state.
Parameters: $a_1=4$, $a_0=0$, $\epsilon=0.01234$, $\delta=1.0.$ }
\label{fig:spiral}
\end{figure}

As a first application of the kinematic equations~(\ref{K})-(\ref{C0})
consider a ``front'' solution connecting the planar Bloch front,
$C_0=C_0^+$, $\kappa=0$, at $s=-\infty$ with the planar Bloch front,
$C_0=C_0^-$, $\kappa=0$, at $s=+\infty$, where $C_0^\pm=
\pm\sqrt{(\alpha_c-\alpha)/\beta}$, and we have assumed a symmetric model,
$a_0=0$ or $\gamma_0=0$. 
Fig.~\ref{fig:spiral}a shows such a solution obtained 
by numerically integrating~(\ref{K})-(\ref{C0}).
As demonstrated in Fig.~\ref{fig:spiral}b this front solution of the kinematic
equations~(\ref{K})-(\ref{C0}) represents a {\em spiral-wave} solution of
the FitzHugh-Nagumo model~(\ref{FHN}).  Unlike the geometrical 
approach~\cite{Mikhailov:90,Meron:92} the spiral core is 
naturally captured by the new kinematic equations.

\begin{figure}[htb]
\epsfxsize=3.5in \epsffile{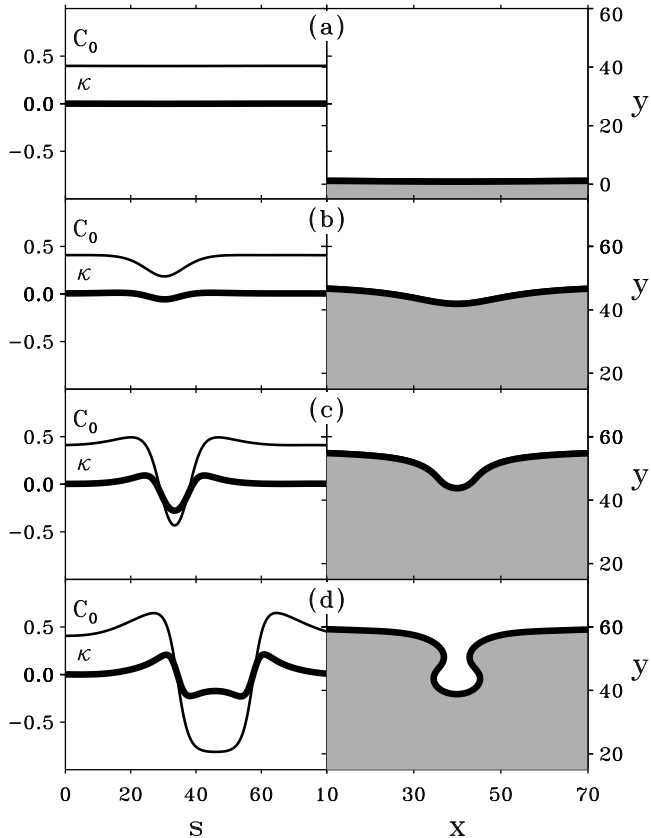}
\caption{Nucleation of a spiral-wave pair in the
kinematic equations  (\protect\ref{K})-(\protect\ref{C0}) 
Left column: the $C_0(s)$ and $\kappa(s)$ profiles. Right column: the front
line shape in the $x-y$ plane.
Parameters: $a_1=4$, $a_0=-0.0001$, $\epsilon=0.0115$, $\delta=1.063.$ 
(a)-(d) are at $t=0,116,136,142$.}
\label{fig:nucleation}
\end{figure}

A second application of the kinematic equations is the study of {\em
spontaneous spiral-wave nucleation}.  Spiral-wave nucleation, induced
by a transverse instability, has been previously observed in direct
simulations of~(\ref{FHN})~\cite{HaMe:94b}.
Figs.~\ref{fig:nucleation}a-d show the time evolution of a solution to
the $C_0-\kappa$ equations representing a planar front near the NIB
bifurcation and beyond the transverse instability boundary.  The
initial front pertains to an up state invading a down state
($C_0>0$). The transverse instability causes a small dent on the front
to grow (Fig.~\ref{fig:nucleation}b). The negative curvature then
triggers the nucleation of a region along the arclength where the
propagation direction is reversed
($C_0<0$)~(Fig.~\ref{fig:nucleation}c). The pair of fronts in the
kinematic equations that bound this region correspond to a pair of
counter-rotating spiral waves in the FitzHugh-Nagumo equations
(Fig.~\ref{fig:nucleation}d).  With this approach, the two-dimensional
spiral-wave nucleation problem is reduced to the considerably simpler
problem of domain, or droplet, nucleation in one
dimension~\cite{Fife:79}.

We have derived kinematic equations for front motion in two-dimensional
bistable systems near a NIB bifurcation. The equations generalize earlier
derivations and capture both the core structure of spiral waves and the
dynamic process of spiral-wave nucleation.  Further investigation is needed
to determine if other features of spiral waves, like the
meander instability~\cite{Meander}, are captured by the equations. Note
that front interaction effects are excluded  by the
choice of the boundary conditions, $v(\mp\infty,s,t)=v_\pm$, in
Eqs.~(\ref{free}).  Such interactions are not significant for the initial
stages of spiral wave nucleation or for the symmetric (or nearly
symmetric) low curvature spirals studied in this Letter.  Front
interactions, however, do become significant when highly curved spirals
develop, and might play an essential role in the meander instability. 
Goldstein {\it et al.}~\cite{GMP:96} have recently studied front
interactions in the fast inhibitor limit ($\epsilon\gg 1$) where
stationary patterns prevail.  A combination of the approaches used in these
two complementary studies may prove useful in establishing a theory of
spiral waves of wider validity. 

%
%

\end{document}